\documentclass[journal]{IEEEtran}
\usepackage[T1]{fontenc}
\usepackage[utf8]{inputenc}
\usepackage[english]{babel}
\usepackage{amsmath,amsfonts, amssymb}
\usepackage{mathtools}
\usepackage{acronym}
\usepackage{longtable}
\usepackage{textcomp}
\usepackage{pifont}
\usepackage{subfigure}
\usepackage{graphicx}
\usepackage{acronym}
\usepackage{color}
\usepackage[linesnumbered]{algorithm2e}
\usepackage{epsfig}
\usepackage{url}
\usepackage{balance}
\usepackage{float}
\usepackage{array}
\usepackage{comment}
\usepackage{pifont}
\usepackage{siunitx}
\usepackage{cite}
\usepackage{enumerate}   
\usepackage{todonotes}

\acrodef{API}{Application Programming Interface}
\acrodef{BPSK}{Binary Phase Shift Keying}
\acrodef{BS}{Base Station}
\acrodef{CDF}{Cumulative Distribution Function}
\acrodef{CP}{Cosine Proximity}
\acrodef{CSS}{Chirp Spread Spectrum}
\acrodef{CSV}{Comma-separated values}
\acrodef{DBPSK}{Differential Binary Phase-Shift Keying}
\acrodef{DBMS}{Database Management System}
\acrodef{ETSI}{European Telecommunications Standards Institute}
\acrodef{GFSK}{Gaussian Frequency Shift Keying}
\acrodef{GPS}{Global Positioning System}
\acrodef{GPU}{Graphic Processing Unit}
\acrodef{HTTP}{HyperText Transfer Protocol}
\acrodef{IPv6}{Internet Protocol Version 6}
\acrodef{IoT}{Internet of Things}
\acrodef{ISM}{Industrial, Scientific and Medical}
\acrodef{JSON}{JavaScript Object Notation}
\acrodef{LoRa}{Long Range}
\acrodef{LPWAN}{Low-Power Wide Area Network}
\acrodef{LSTM}{Long Short-Term Memory}
\acrodef{LTE}{Long-Term Evolution}
\acrodef{MAC}{Medium Access Control}
\acrodef{MAE}{Mean Absolute Error}
\acrodef{MCU}{MicroController Unit}
\acrodef{MQTT}{MQ Telemetry Transport}
\acrodef{MSE}{Mean Square Error}
\acrodef{NB-IoT}{Narrow-Band IoT}
\acrodef{OTA}{Over-the-Air}
\acrodef{PCA}{Principal Component Analysis}
\acrodef{PER}{Packet Error Rate}
\acrodef{PHY}{PHYsical layer}
\acrodef{QPSK}{Quadrature Phase Shift Keying}
\acrodef{RAN}{Radio Access Network}
\acrodef{ROM}{Read Only Memory}
\acrodef{ReLU}{Rectified Linear Unit}
\acrodef{RF}{Radio Frequency}
\acrodef{RFTDMA}{Random Frequency and Time Division Multiple Access}
\acrodef{RNN}{Recurrent Neural Network}
\acrodef{SNMP}{Simple Network Management Protocol}
\acrodef{UNB}{Ultra Narrow Band}
\acrodef{WHO}{World Health Organization}
\acrodef{WSN}{Wireless Sensor Network}

\newcolumntype{P}[1]{>{\centering\arraybackslash}p{#1}}

\newcommand{\cmark}{\ding{51}}
\newcommand{\xmark}{\ding{53}}

\graphicspath{{figures/}}
\begin{document}
\title{Water Quality Prediction on a Sigfox-compliant IoT Device: The Road Ahead of WaterS}

\author{\IEEEauthorblockN{Pietro Boccadoro\IEEEauthorrefmark{1}\IEEEauthorrefmark{3} \textit{IEEE Student Member}, Vitanio Daniele\IEEEauthorrefmark{5}, Pietro Di Gennaro\IEEEauthorrefmark{6}, Domenico Lofù\IEEEauthorrefmark{1}\IEEEauthorrefmark{4} \textit{IEEE Student Member}, Pietro Tedeschi\IEEEauthorrefmark{2} \textit{IEEE Student Member}\\}
    
    \IEEEauthorblockA{\IEEEauthorrefmark{1}Dept. of Electrical and Information Engineering (DEI), Politecnico di Bari, Bari (Italy)\\\{pietro.boccadoro, domenico.lofu\}@poliba.it\\}
    \IEEEauthorblockA{\IEEEauthorrefmark{2}Division of Information and Computing Technology (ICT), College of Science and Engineering (CSE), Hamad Bin Khalifa University (HBKU), Doha (Qatar) \\ptedeschi@hbku.edu.qa\\}
    \IEEEauthorblockA{\IEEEauthorrefmark{3}CNIT, Consorzio Nazionale Interuniversitario per le Telecomunicazioni, Politecnico di Bari, Bari (Italy)\\}
    \IEEEauthorblockA{\IEEEauthorrefmark{4}Innovation Lab, Exprivia S.p.A., Molfetta (Italy) \\domenico.lofu@exprivia.com\\}
    \IEEEauthorblockA{\IEEEauthorrefmark{5}Connected Vehicle \& Micro Mobility, Sitael S.p.A., Mola di Bari (Italy)\\vitanio.daniele@sitael.com\\}
    \IEEEauthorblockA{\IEEEauthorrefmark{6}Fincons Group S.p.A., Bari (Italy)\\pietro.digennaro@finconsgroup.com}
    }
\maketitle

\begin{abstract}
Water pollution is a critical issue that can affects humans' health and the entire ecosystem thus inducing economical and social concerns. In this paper, we focus on an \acl{IoT} water quality prediction system, namely WaterS, that can remotely communicate the gathered measurements leveraging \acl{LPWAN} technologies. The solution addresses the water pollution problem while taking into account the peculiar \acl{IoT} constraints such as energy efficiency and autonomy as the platform is equipped with a photovoltaic cell.
At the base of our solution, there is a \acl{LSTM} recurrent neural network used for time series prediction. It results as an efficient solution to predict water quality parameters such as pH, conductivity, oxygen, and temperature. The water quality parameters measurements involved in this work are referred to the \textit{Tiziano Project} dataset in a reference time period spanning from 2007 to 2012. The LSTM applied to predict the water quality parameters achieves high accuracy and a low \acl{MAE} of $~0.20$, a \acl{MSE} of $~0.092$, and finally a \acl{CP} of $~0.94$.
The obtained results were widely analyzed in terms of protocol suitability and network scalability of the current architecture towards large-scale deployments. From a networking perspective, with an increasing number of Sigfox-enabling end-devices, the \acl{PER} increases as well up to $4$\% with the largest envisioned deployment.
Finally, the source code of WaterS ecosystem has been released as open-source, to encourage and promote research activities from both Industry and Academia.
\end{abstract}

\IEEEpeerreviewmaketitle

\section{Introduction}\label{sec:intro}
The \ac{IoT} is a well-known paradigm that turns devices into interconnected smarter objects. \ac{IoT} devices are generally characterized by low computational power, networking limitations, and communication capabilities. As a matter of fact, \ac{IoT} devices have to deal with issues related to data exchange while optimizing the communication protocols in terms of latencies, bandwidth, security, and energy consumption~\cite{Lin2017_IoTJ, Tedeschi2020_IoTJ, WSS2018}. Despite their intrinsic limitations, \ac{IoT} devices are now part of continuous monitoring processes in several fields, from industrial applications~\cite{Sisinni2018_TII}, to monitoring activities connected to air quality and environmental parameters in the most modern smart cities~\cite{Montori2018_IoTJ}.

Among the many fields that could benefit from the introduction of \ac{IoT} technologies, water quality monitoring is certainly one of the most relevant and recently investigated \cite{Myint2017_ICIS, Pranata2017_LANMAN, Cloete2016_IEEEA, Kamaludin2017_ICSPC, Anand2018_ICCICT, Sammoudi2019_SCA, Wang2018_ICCET, Liu2019_SUISTAINABILITY, Mukta2019_ICCCS}.
In this context, WaterS~\cite{DLV+19} has been already proven to be able to provide remote monitoring capabilities for some of the most representative water quality indicators (i.e., temperature and turbidity).
At the same time, the WaterS architecture provides an energy-harvesting and an ultra-low-power Sigfox-compliant radio interface to keep track of the continuous monitoring activities carried out by the \ac{IoT} architecture.
Even though the monitoring activities are of utmost importance to grant fine-grained sampling of the parameters of interest, water pollution, and other long-lasting phenomena, could be still not detected.
In fact, since a large portion of the world's freshwater lies underground, infiltration into the ground could be underestimated.
Therefore, a system like WaterS could be more and more useful if it was able to carry out a prediction analysis on water quality parameters.
Such an advancement could lead to the massive adoption of smart and energy-efficient sensing units to be employed in hostile areas, for example, subject to massive and pervasive pollution phenomena such as the spillage of toxic waste into the aquifers where water infiltration is a crucial task~\cite{Richardson2018_AC}.

To achieve the ambitious goals of measuring and forecasting water quality parameters, the proposed system is stand-alone, energy-efficient, and standard-compliant.

The system was tested close to the seaside in the city town of Bari, Italy. The dataset involved in this study is one of the main outcomes of the Tiziano project~\cite{progettoTiziano} and it has been fully exploited to train a deep learning model for forecasting, i.e., in our case a \ac{LSTM} neural network. WaterS has been developed by adopting open-source hardware/software (with a focus on the energy harvesting \cite{tedeschi2020security} capabilities of our solution) and a standardized wireless protocol to further push the innovation, as well as to allow researchers and academia to further use our code as a ready-to-use basis for further software development~\cite{code}.

\textbf{Contribution.} This work aims at integrating an advanced deep learning technique namely \ac{LSTM} within WaterS to improve the current solution by providing additional features like the water quality prediction.
Specifically, we provide an experimental evaluation where we show how the adoption of \ac{LSTM} can effectively predict the reference data by assessing the results accordingly the \acl{MSE}, \acl{MAE}, and \acl{CP} metrics.
In particular, the neural network has been configured to search for correlations on multivariate time series on surface water. Indeed, the experimental results demonstrated that some degree of correlation exists and this proves that it is worth pursuing estimations on the proposed variables. In addition, the achieved results show that the adoption of \ac{RNN} for the analysis of water quality is a winning solution for the study of multivariate time series~\cite{Zhang2018_JoH}.
Comparisons against competing solutions show the viability and efficiency of our proposal. Finally, WaterS has been fully implemented as the first open hardware/software solution, and the source code has been released as open-source~\cite{code}. This permits the research community and companies to reproduce our results, use the solution on top of existing Sigfox transceivers, adopt the released code as a ready-to-use basis for further improvements and comparison and, finally, allow the interested readers to verify our claims.

\textbf{Roadmap.} The remainder of the present work is as follows: Section~\ref{sec:background_related} is three-folded, since it introduces the reference background on (i) water quality monitoring \ac{IoT} systems, (ii) \acp{LPWAN} technologies, with a focus on the Sigfox protocol, and (iii) a thorough analysis on \ac{LSTM}. Section~\ref{sec:waters_deep} describes the operating scenario in which WaterS is adopted, as well as the envisioned architecture and the proposed prediction system. 
Section \ref{sec:approach} summarises both the leading criteria and methodological approach. On top of that, Section~\ref{sec:experiments} presents the experimental campaign while Section \ref{sec:results} discusses the obtained results. Possible strategies for improving the WaterS systems are proposed in Section~\ref{sec:disc-way-forw} together with the main findings, limitations, and future research directions. Finally, Section~\ref{sec:conclusion} tightens the conclusions.

\section{Background and Related Work} \label{sec:background_related}
This section provides the background on \ac{IoT} systems specifically designed for environmental monitoring activities. In general, some of them are focused on water quality control, whereas some others are devoted to air quality. In almost all of them, one of the key features is the ability to communicate with remote users/base stations. This data-gathering activity usually enables advanced analysis possibilities.
The largest majority of the surveyed contribution deals with \ac{LPWAN} communications, leveraging some of the newest standardized solutions/protocols, with a focus on Sigfox.
Since the \ac{IoT} domains/applications are usually aimed at improving sensing, elaboration, and communications, what data can be used for is still left unspoken.
Nevertheless, the data-gathering phase can be considered as a prerequisite for the analysis of simple and/or complex phenomena. In this context, machine learning techniques can be successfully involved. In this regard, a deep learning solution based on \ac{LSTM} will be discussed with a focus on the analysis of multivariate datasets with specific reference to water quality testing, monitoring, and predicting capabilities.

\subsection{Related Work}\label{sec:related_work}
The employment of machine learning has been recently proposed for several applications and research fields, such as environmental monitoring, smart grid, water treatment facilities, and power plants \cite{wang2017evaluation,zhang2018lstm, Wang2019_IoTJ}, to name a few. Some of these contributions consider also the use of deep learning, a subset of machine learning that allows unsupervised learning. The main difference between these two relies on the way data is presented in the system. In fact, unlike machine learning, deep learning does not need structured input data. So, human intervention is not always mandatory, as multilevel layers of artificial neural networks are able to automatically recognize common features and learn from data without external help.

Manu \emph{et al.} \cite{Manu2017_TSN} provide an overview of the design and implementation of water quality monitoring systems. Furthermore, it discusses wireless technologies and their adoption at each stage of the monitoring processes.
Sammoudi \emph{et al.} \cite{Sammoudi2019_SCA} characterized a water quality assessment system through the analysis of physicochemical and bacteriological water properties. Furthermore, the authors adopted the \ac{PCA} technique to identify potential indicators for water quality measurement.
Liu \emph{et al.} \cite{Liu2019_SUISTAINABILITY} introduced a water quality prediction model leveraging \ac{LSTM} deep neural networks. They collected data by using third-party water monitoring stations. The results demonstrated that the model can predict water quality over a $6$ months period. Even though the results are of relevance, the main limitations of the proposed prediction model are related to 1-dimensional inputs and limited data-set.
All in all, the surveyed state of the are demonstrated that those limitations are highlighted by many contributions, thus suggesting that the method itself benefits from multi-dimensional inputs and huge datasets. 

The research activities carried out on the theme also had business counterparts, with many different industrial-grade solutions. In particular, those solutions took into account diversified sensing units, precisions, and capabilities to provide detailed analysis on peculiar water parameters (e.g., pH, water level, turbidity, carbon dioxide, and temperature) \cite{Myint2017_ICIS, Pranata2017_LANMAN, Cloete2016_IEEEA}. In these solutions, acquired data are remotely transmitted thanks to some of the main short-range \ac{IoT} protocol (i.e., Zigbee) and stored within the server that received the messages to be investigated and elaborated, as needed.
Kamaludin \textit{et al.} \cite{Kamaludin2017_ICSPC} discussed the implementation of water monitoring \ac{IoT} system platform for water quality monitoring \ac{IoT} system working in the sub-GHz bands to gather the pH of waters as well as environmental temperature values.
Wang \emph{et al.} \cite{Wang2018_ICCET} envisaged a LoRa based \ac{IoT} system that leverages ultrasonic water meters, data centralizers, and intelligent remote valves to detect and predict system leaks and prevent water theft.
In \cite{Anand2018_ICCICT}, a remote water level monitoring system is proposed. Gathered data communications are enabled thanks to the employment of NB-IoT, a choice motivated by the fact that narrowband technology proposes several advantages in terms of optimized data rate and enlarged coverage area.
Mukta \emph{et al.} \cite{Mukta2019_ICCCS}, instead, proposed the development of an \ac{IoT} system specifically designed for monitoring the physical parameters of drinking water. In this case, the system is used to measure temperature, pH, electrical conductivity, and turbidity locally. The system is also in charge of using a binary classifier to state the quality of the water.

All the previously introduced solutions were designed to address water quality monitoring/detection, thanks to peculiar parameters.
Nevertheless, despite the importance of the surveyed solutions, none of them was able to thoroughly fulfill the key requirements and challenges of the \ac{IoT} domain.
Counterwise, the WaterS system \cite{DLV+19} proposes itself as a thorough, stand-alone, energy-efficient, and standard-compliant prototype. It envisages a real experimental testbed for measuring and forecasting water quality parameters. Table~\ref{tab:comparison} summarises the main findings.

\begin{table*}[htbp]
\centering
    \caption{Comparison between concepts, devices, systems and prototypes for Water quality measurements and analysis. A \checkmark\ symbol indicates the fulfillment of a particular feature, a \xmark\ symbol denotes the miss of the feature or that the feature is not applicable.}
    \begin{tabular}{|l||c|c|c|c|c|c|c|c|c|c|}
    \hline
        \textbf{Feature} & \cite{Myint2017_ICIS} & \cite{Pranata2017_LANMAN} & \cite{Cloete2016_IEEEA} & \cite{Kamaludin2017_ICSPC} & \cite{Anand2018_ICCICT} & \cite{Sammoudi2019_SCA} & \cite{Wang2018_ICCET} & \cite{Liu2019_SUISTAINABILITY} & \cite{Mukta2019_ICCCS} & \textbf{WaterS}\\ \hline\hline
        \emph{Preliminary mathematical formulation} & \xmark & \xmark & \xmark & \xmark & \xmark & \checkmark & \xmark & \checkmark & \xmark & \cmark\\ \hline
        \emph{Simulation Validation} & \checkmark & \checkmark & \checkmark & \checkmark & \checkmark & \checkmark & \checkmark & \checkmark & \checkmark & \cmark\\ \hline
        \emph{Experimental Validation} & \xmark & \checkmark & \checkmark & \checkmark & \xmark & \checkmark & \checkmark & \xmark & \checkmark & \cmark\\ \hline
        \emph{Open Source Hardware} & \checkmark & \checkmark & \checkmark & \checkmark & \xmark & \xmark & \xmark & \xmark & \checkmark & \cmark\\ \hline
        \emph{Open Source Code} & \xmark & \xmark & \xmark & \xmark & \xmark & \xmark & \xmark & \xmark  & \xmark & \cmark\\ \hline
        \emph{Energy Harvesting} & \xmark & \xmark & \xmark & \xmark & \xmark & \xmark & \xmark & \xmark & \xmark & \cmark\\ \hline
        \emph{\ac{RF} communications} & \checkmark & \checkmark & \checkmark & \checkmark & \checkmark  & \xmark  & \checkmark & \xmark & \xmark  & \cmark\\ \hline
        \emph{Standard-compliance} & \checkmark & \checkmark & \checkmark & \checkmark & \checkmark  & \xmark & \checkmark & \xmark & \xmark & \cmark\\ \hline
        \emph{Cloud-based monitoring} & \xmark & \xmark & \xmark & \checkmark  & \xmark & \xmark & \checkmark & \xmark & \xmark & \cmark\\ \hline
        \emph{Available \acp{API}} & \xmark & \xmark & \xmark & \xmark & \xmark & \xmark & \xmark & \xmark & \xmark & \cmark\\ \hline
        \emph{Prediction Capabilities with Machine Learning} & \xmark & \xmark & \xmark & \xmark & \xmark & \xmark & \checkmark & \checkmark & \checkmark & \cmark\\ \hline
    \end{tabular}
\label{tab:comparison}
\end{table*}

\subsection{The Sigfox technology} \label{sec:sigfox}
\acp{LPWAN} are widely conceived as the winning solution in multiple \ac{IoT} applications/domain \cite{Al-Fuqaha_2015_Comst}. This is motivated by the fact that those technologies jointly optimize the amount of transmitted data while maximizing coverage area.
\acp{LPWAN} are taken into consideration because of the enhancements they propose, such as the friendly up-scaling. It is worth noting that these advantages are achievable while lowering the energy footprint required by \ac{IoT} networks.

\begin{figure}[!ht]
    \centering
    \includegraphics[angle=0, width=\columnwidth]{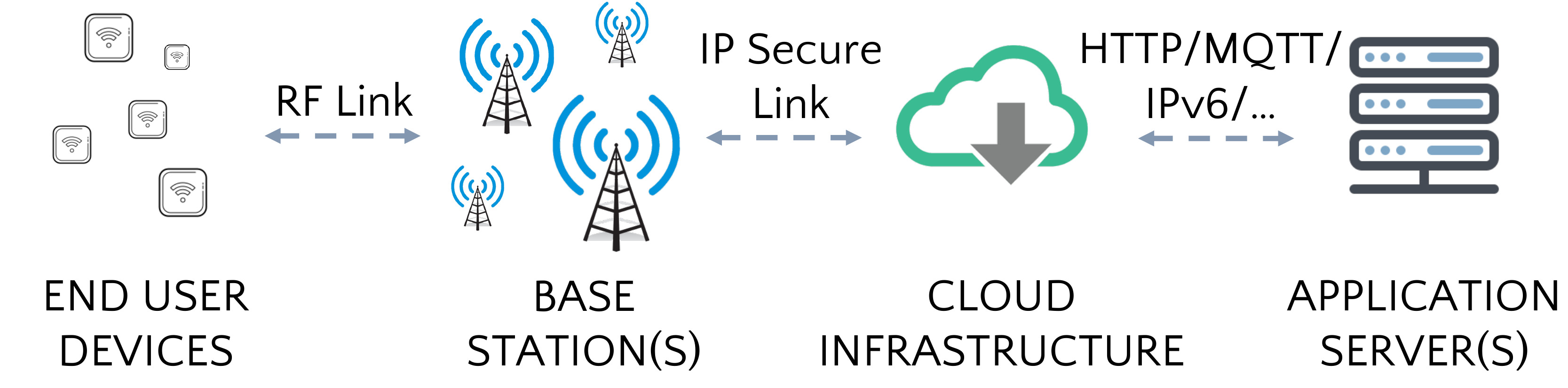}
    \caption{Sigfox Network Architecture.}
    \label{fig:sigfox_net_arch}
\end{figure}

\begin{table*}[htbp]
\centering
\caption{Comparison of the main features of the \ac{LPWAN} technologies}
    \begin{tabular}{l|l|l|l|}
    \cline{2-4}
                                                        & \multicolumn{1}{c|}{\textbf{Sigfox}}                                  & \multicolumn{1}{c|}{\textbf{LoRa}}                                   & \multicolumn{1}{c|}{\textbf{NB-IoT}}                                 \\ \hline
    \multicolumn{1}{|l|}{\emph{Modulation}}           & \ac{BPSK}                                                                  & \ac{CSS}                                                                  & \ac{QPSK}                                                                 \\ \hline
    \multicolumn{1}{|l|}{\emph{Frequency}}            & Unlicensed ISM bands sub-GHz                                                  & Unlicensed ISM bands sub-GHz                                                 & Licensed \ac{LTE}                                                         \\ \hline
    \multicolumn{1}{|l|}{\emph{Bandwidth}}            & $100$~Hz                                                                & $125, 250, 500$~kHz                                                    & $200$~kHz                                                              \\ \hline
    \multicolumn{1}{|l|}{\emph{Network Topology}}     & Star                                                                  & Star on Star                                                         & Star                                                                  \\ \hline
    \multicolumn{1}{|l|}{\emph{DataRate}}             & $100$~bps                                                               & $50$~kbps                                                              & $200$~kbps                                                             \\ \hline
    \multicolumn{1}{|l|}{\emph{Message/day (MAX)}}    & $140$ (UL), $4$ (DL)                                                      & Unlimited                                                            & Unlimited                                                            \\ \hline
    \multicolumn{1}{|l|}{\emph{Payload length (MAX)}} & $12$~B (UL), $8$~B (DL)                                                   & $243$~B                                                                & $1.6$~kB                                                               \\ \hline
    \multicolumn{1}{|l|}{\emph{Range}}                & \begin{tabular}[c]{@{}l@{}}$10$~km (Urban)\\ $40$~km (Rural)\end{tabular} & \begin{tabular}[c]{@{}l@{}}$5$~km (Urban)\\ $20$~km (Rural)\end{tabular} & \begin{tabular}[c]{@{}l@{}}$1$~km (Urban)\\ $10$~km (Rural)\end{tabular} \\ \hline
    \multicolumn{1}{|l|}{\emph{End-node Roaming}}     & Yes                                                                   & Yes                                                                  & Yes                                                                   \\ \hline
    \multicolumn{1}{|l|}{\emph{Licensed use}}         & No                                                                   & No                                                                  & Yes                                                                   \\ \hline

    \end{tabular}
    \label{tab:lpwans_comparison}
\end{table*}
Sigfox is a narrowband \ac{LPWAN} proprietary protocol that uses the unlicensed \ac{ISM} frequencies bands~\cite{Al-Fuqaha_2015_Comst}. Compared with \ac{LoRa} and \ac{NB-IoT}, Sigfox allows minimizing the exchange of packets, reduce the bandwidth consumption, also limiting the energy consumption during radio communications. In Table \ref{tab:lpwans_comparison}, a summary of the main features and functional features of \ac{LPWAN} technologies is given. Nowadays, Sigfox technology enables several \ac{IoT} applications, such as remote tracking, smart parking, waste management, environmental monitoring, real-time health, and fitness monitoring, and public safety to name a few. As depicted in Figure~\ref{fig:sigfox_net_arch} the standard Sigfox network architecture envisages four layers: (i) \emph{end user devices}, (ii) one or more Sigfox \emph{base station(s)}, (iii) the Sigfox \emph{cloud architecture} and finally (iv) the \emph{application server(s)}. The end devices are connected with the gateway in a star topology, by leveraging radio frequency links. Besides, there is a secure link between the base station(s) and the cloud infrastructure. Finally, the communication between the cloud architecture and the application server(s) can be established leveraging different protocols such as \ac{SNMP}, \ac{MQTT}, and \ac{HTTP}.

The Sigfox standard supports up to $140$ uplink messages a day (duty cycle of $1$\%, $6$ messages/hour), each with an uplink payload of $12$~bytes and an $8$~bytes one in downlink. The data rate goes up to $100$~bps in uplink and $600$~bps in downlink. Each data transmission requires approximately $6$~seconds. The Sigfox protocol stack consists of different layers described as follows.

\textbf{Physical Layer.} Sigfox technology is an \ac{UNB} deployed over a bandwidth of $192$~kHz with each transmission $100$~Hz wide. According to the \ac{ETSI} $300-220$ regulation, in Europe, the frequency band adopted is $868$~MHz, while in North America according to the FCC part 15 regulations is $902$~MHz. Uplink messages are modulated with \ac{DBPSK}, while downlink messages are modulated with a \ac{GFSK}. The adoption of these modulation schemes, enables the devices to communicate in a range between $10$ to $50$~km with low power consumption (the maximum uplink and downlink transmission power is set to $25$~mW and $500$~mW in Europe, while it is set to $158$~mW and $4$~W in the USA).

\begin{figure}[!ht]
    \centering
    \includegraphics[angle=0, width=\columnwidth]{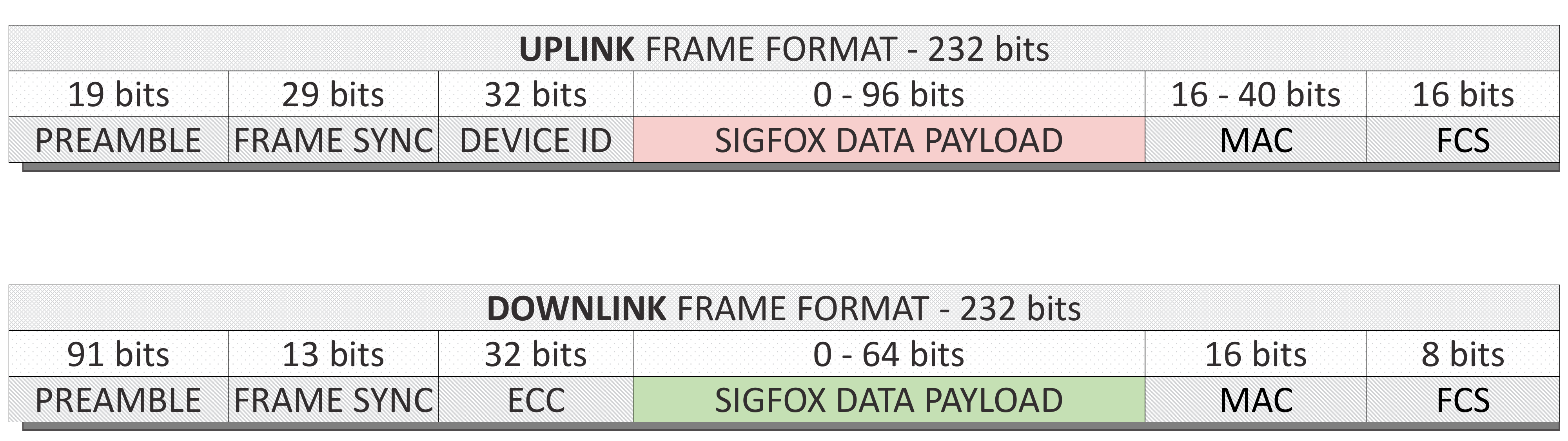}
    \caption{Sigfox Frame Structure.}
    \label{fig:sigofx_frame}
\end{figure}

\textbf{MAC Layer.} The Sigfox MAC layer is based on the unslotted Aloha MAC protocol. Access to the wireless medium channel relies on \ac{RFTDMA}. According to the standard Sigfox, each message can be sent up to $3$ times on different frequencies to improve reliability. As shown in Figure~\ref{fig:sigofx_frame}, Sigfox uplink frames have an overall size of $232$~bits, while Sigfox downlink frames have an overall size of $224$~bits. The uplink frame starts with a \emph{preamble} of $19$~bits of predefined symbols, used to identify an upcoming Sigfox message and from the receiver side to synchronize with the symbols sent by the transmitter. The \emph{frame synchronization} field of $29$~bits specifies the type of the frames to be transmitted, while the \emph{end-device-id} of $32$~bits is an unique identifier for each Sigfox device that is adopted for routing and signing frames. The \emph{payload} field ranges up to $96$~bits are devoted to the data, while the \emph{Message Authentication Code} that spans from $16$ to $40$~bits and provides the frame authenticity. Finally, the last $16$~bits identify the \emph{Frame Check Sequence} intending to detect communication errors. On the other hand, the downlink frame starts with a \emph{preamble} of $91$~bits and the \emph{frame synchronization} field of $13$~bits. The \emph{Error-Correcting-Code} of $32$~bits is used to detect errors in the data payload, and finally the \emph{payload} field up to $64$~bits, the \emph{Message Authentication Code} of $16$~bits and the \emph{Frame Check Sequence} of $16$~bits are used in the same way as specified for the uplink data structure \cite{zuniga2016sigfox}.

\textbf{Frame Layer.} This layer allows the generation of the radio frames starting from the \emph{Application Layer}. Further, it attaches a sequence number during data transmission.

\textbf{Application Layer}. This layer is devoted to managing functionality like messaging and web-services.

In terms of security, Sigfox frames are not encrypted by design. In detail, the (i) confidentiality is provided at the application layer, the (ii) authenticity is provided by the Message Authentication Codes, and finally, the (iii) replay attacks prevention is provided by the sequence-number defined in the message frame.

\subsection{\ac{LSTM}} \label{sec:lstm}
\ac{LSTM} is an example of \acp{RNN} proposed for deep learning by Hochreiter and Schmidhuber~\cite{Hochreiter1997}. The main feature of the \ac{LSTM} is defined by the feedback connections, that are used to store representations of recent input events in the form of activations. As demonstrated in the scientific literature, these types of networks allow to effectively prevent the gradient vanishing and explosion problems during back-propagation through time by keeping the error constant during the learning phase. An \ac{LSTM} network is built by leveraging multiple \ac{LSTM} cells. The notation to describe this network is summarized as reported in Table~\ref{tab:lstm_notation}.
\begin{table}[htbp]
    \caption{Notation used for the LSTM description.}
    \centering
        \begin{tabular}{c|c}
        \textbf{Notation} & \textbf{Description} \\ \hline
        $t$ & Time Step \\ \hline
        $x_t$ & Input Vector \\ \hline
        $f_t$ & Forgetting Gate \\ \hline
        $i_t$ & Input/Update Gate \\ \hline
        $o_t$ & Output Gate \\ \hline
        $h_t$ & Hidden State Vector  \\ \hline
        $C_t$ & Memory Cell \\ \hline
        $\tilde{C_t}$ & Candidate State \\ \hline
        $\sigma$ & Sigmoid Function \\ \hline
        $tanh$ & Hyperbolic Tangent Function \\ \hline
        $\mathbf{W}_f,\mathbf{W}_i,\mathbf{W}_c,\mathbf{W}_o$ & Weight Matrices \\ \hline
        $b_f,b_i,b_c,b_o$ & Bias Vectors \\ \hline
        $\circ$ & Hadamard Product Operator \\ \hline
        \end{tabular}
    \label{tab:lstm_notation}
\end{table}
\begin{figure}[htbp]
  \centering
  \includegraphics[angle=0, width=\columnwidth]{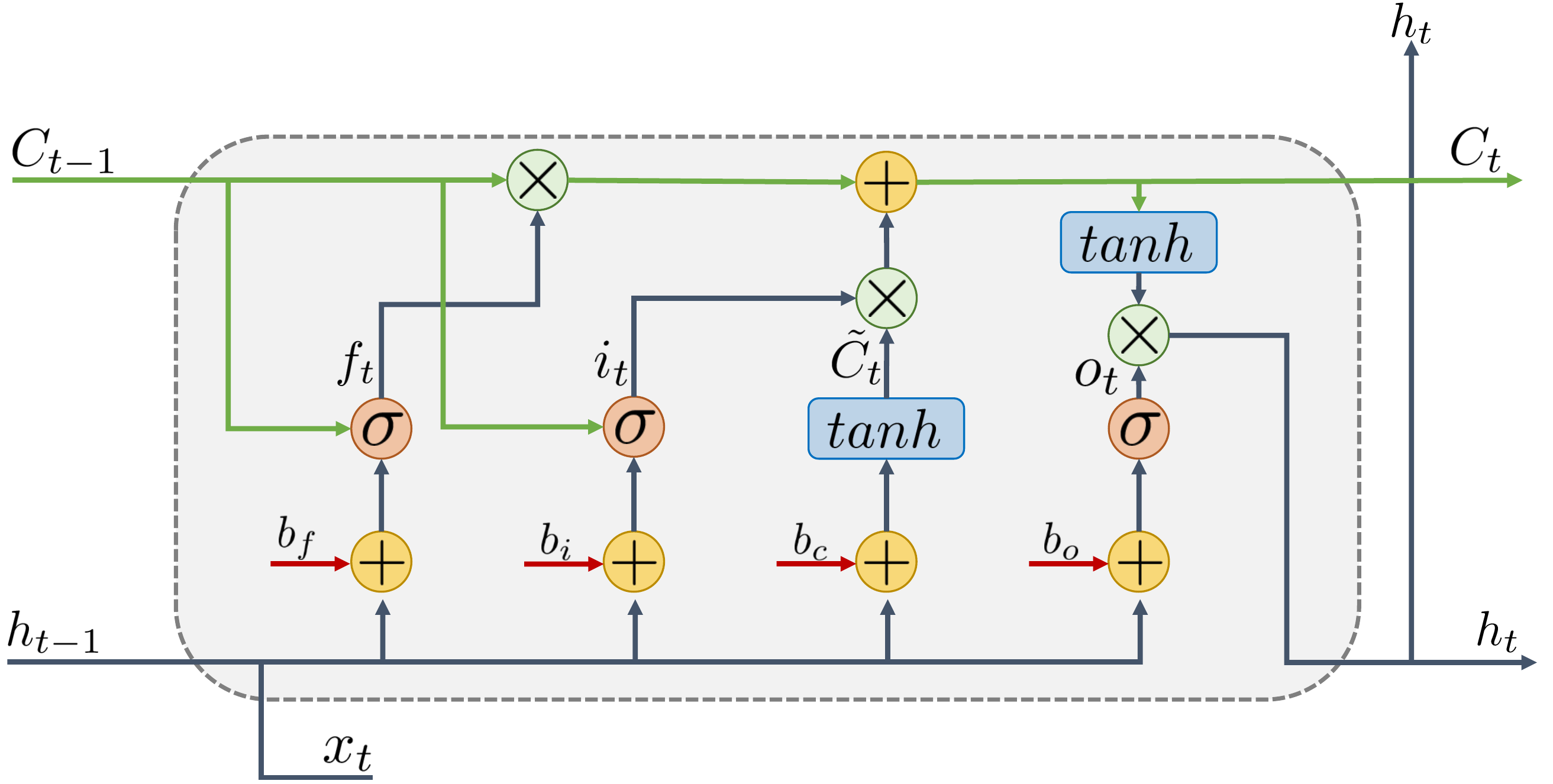}
  \caption{A diagram of a standard \ac{LSTM} memory cell.}
  \label{fig:lstm}
\end{figure}

The \ac{LSTM} architecture can be defined by the following equations:
\begin{align}
    \label{eq:ft}
    f_t&=\sigma(\mathbf{W}_f\cdot[h_{t-1},x_t]+b_f) \\
    \label{eq:it}
    i_t&=\sigma(\mathbf{W}_i\cdot[h_{t-1},x_t]+b_i) \\
    \label{eq:ctilde}
    \tilde{C_t}&=tanh(\mathbf{W}_c\cdot[h_{t-1},x_t]+b_c)\\
    \label{eq:ct}
    C_t&=f_t\circ C_{t-1}+i_t\circ \tilde{C_t}\\
    \label{eq:ot}
    o_t&=\sigma(\mathbf{W}_o\cdot[h_{t-1},x_t]+b_o) \\
    \label{eq:ht}
    h_t&=o_t\circ tanh(C_t)
\end{align}
As shown in the Figure~\ref{fig:lstm}, an \ac{LSTM} network adopts memory units in order to: (i) learn and store values over arbitrary temporal intervals, (ii) forget the previously hidden states, and (iii) update the hidden states with new data. The input and output information flows for each \ac{LSTM} cell is controlled by an \emph{input gate}, an \emph{output gate}, and a \emph{forgetting gate}.

When data are provided as input to the \ac{LSTM}, the first operational step consists of establishing which information must be kept in memory and which must be discarded. This decision is made in accordance to the mathematical formulation proposed in Eq.~\eqref{eq:ft}, which takes as input (i) the vector $x_t$ at the generic instant of time $t$ and (ii) the previous output $h_{t-1}$ at time $t-1$. The computed value is given as input to the sigmoid function $\sigma$ which returns a number between $0$ and $1$ for each element of the state cell $C_{t-1}$, where $0$ means that the element can be discarded and $1$ means that the element must be maintained over time.

According to Eq.~\eqref{eq:it}, in the second step new information is transferred through the input gate. The obtained output value is between $0$ and $1$, where $1$ means that the information must be updated, and $0$ means that it can be discarded.

The third step defines the state of the memory cell $C_t$ at the time $t$. In Eq.~\eqref{eq:ctilde}, the hyperbolic tangent $tanh$ is envisioned as the activation function adopted to normalize the output between $-1$ and $1$. The choice is motivated by the fact that the problem of the vanishing gradient for \acp{RNN} must be solved \cite{Jozefowicz2015}.

Finally, the new state of the \ac{LSTM} is updated through Eq.~\eqref{eq:ct} (initially set to zero). It is a combination of values computed at the instant of the current time $t$ and previous time $t-1$. The output generated by the \ac{LSTM} is defined by Eq.~\eqref{eq:ot} and Eq.~\eqref{eq:ht} (initially set to zero)~\cite{Greff2017}. It is worth noting that the weight matrices $\mathbf{W}_j, j \in \{c,f,i,o\}$ and the bias vectors $\mathbf{b}_j, j \in \{c,f,i,o\}$ are optimization parameters for the \ac{LSTM}.

\section{Design and System Model}\label{sec:waters_deep}
The background and the requirement analysis carried out in the previous section allow us to describes the operating scenario in which the WaterS system is at work, as well as the envisioned architecture and the proposed deep learning solution.

According to the conditions of the environment in which the surveys are carried out, all the water quality parameters may be subject to significant changes over time. One of the most important aspects of this contribution consists of improving the current proposal with deep learning capabilities. Indeed we investigated, how unexpected changes in these values, can prove water pollution phenomena. On top of this evaluation, WaterS can exploit what has been learned to make future predictions in the same site, or different sites with natural overlapping water conditions.

\begin{figure}[!ht]
    \centering
    \includegraphics[angle=0, width=\columnwidth]{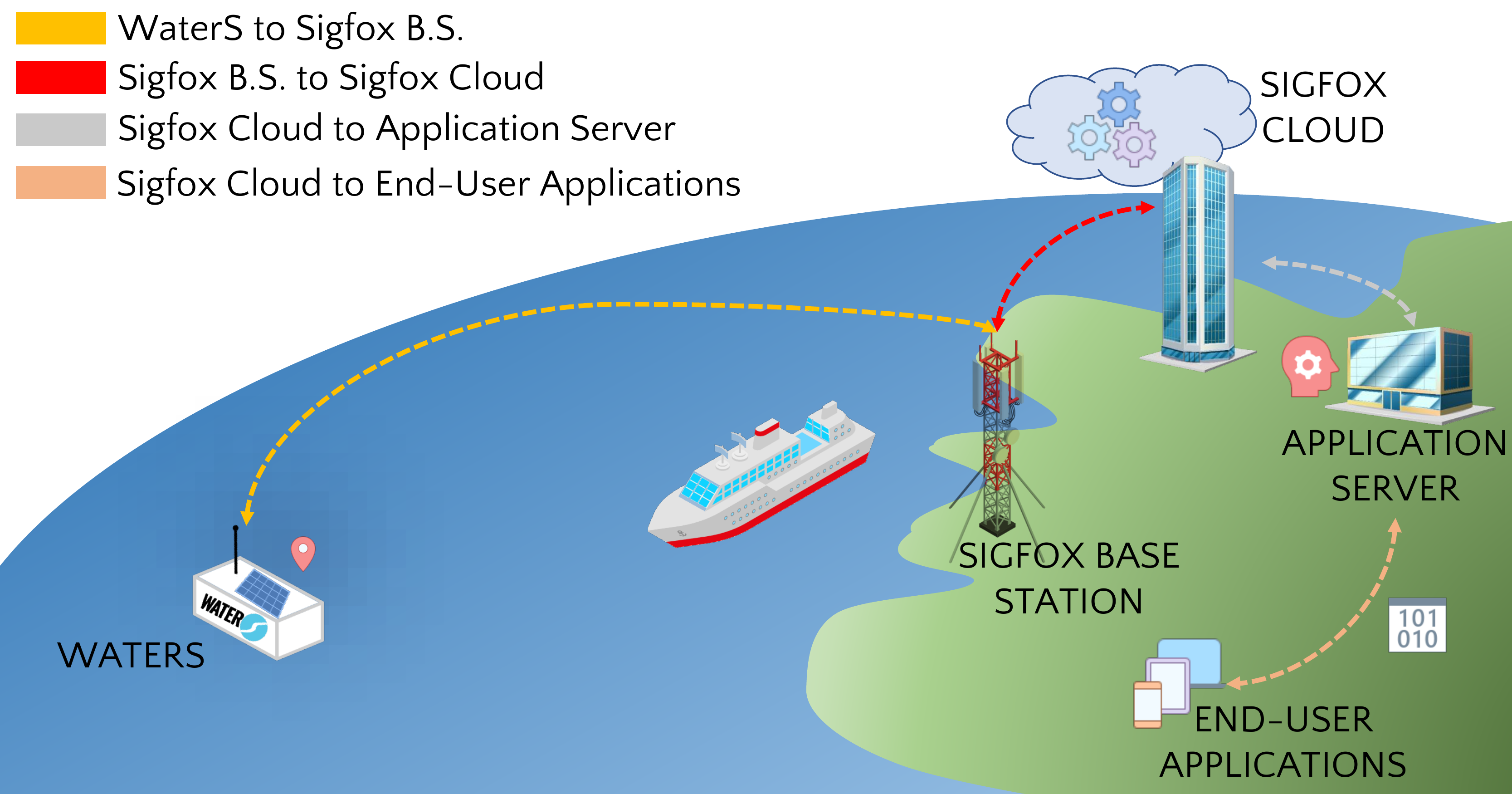}
    \caption{Operating Scenario.}
    \label{fig:scenario}
\end{figure}

The operating scenario includes the WaterS end-device prototype carrying out monitoring activities conveyed out in open waters, as depicted in Figure~\ref{fig:scenario}.

\textbf{The WaterS system.} The system architecture assumed by WaterS involves the following entities: (i) the \ac{IoT} end-node, i.e. our WaterS prototype, (ii) the \emph{Sigfox \ac{BS}}, (iii) the \emph{Sigfox Cloud}, (iv) the \emph{application server}, and (v) an end-user application.
As for the former, the WaterS~\cite{DLV+19} end-node can be defined as an \ac{IoT} device mainly conceived as an energy-efficient and Sigfox-compliant sensing unit, able to periodically gather geo-referenced water quality information.
The sensed values are properly handled and pre-processed by the remote network server according to the Sigfox protocol.

WaterS is made up of: (a) the mainboard, (b) the sensing units such as a pH probe, a turbidity sensor, and a thermal probe, and (c) a UBLOX NEO‐8M \ac{GPS} module to enable the data geolocation. The mainboard used to acquire and process all the sampled data from the aforementioned sensors is an Arduino MKRFOX1200, based on the Microchip SAMD21 Micro Controller Unit with a $48$~MHz clock speed, and an ATA8520 Sigfox module.
The autonomy of the WaterS end-device is guaranteed thanks to two power sources: a $3.7$ V‐$720$~mAh LiPo battery, that is the main power supply, and a solar shield, suitable to increase the energy budget required to power the device and then to address the energy availability and consumption critical issues.

According to the detailed analysis on energy consumption in Table~\ref{tab:consumption_specs}, the \ac{IoT} device has been characterized in terms of autonomy thus deriving a total of $18$~hours, assuming a previous $6$ hours period of full daylight, which implies ideal conditions for a full recharge of the battery supply the WaterS end-device is equipped with.

\begin{table}[htbp]
    \caption{Consumption specifications for each involved component.}
    \centering
    \begin{tabular}{c|c|c}
        &\textbf{Wake Mode (mA)}&\textbf{Sleep Mode (mA)}   \\ \hline
        MCU Antenna & $6.48$ & $0.0043$                     \\ \hline
        PH probe & $10.0$ & $0.28$                          \\ \hline
        Turbidity & $40.0$ & $0.4$                          \\ \hline
        Temperature probe & $2.0$ & $0.1$                   \\ \hline
        GPS Antenna& $67.0$ & $0.2$                         \\ \hline
        Total & $125.5$ & $0.98$                            \\ \hline
     \end{tabular}
     \label{tab:consumption_specs}
\end{table}
The prototype carries out a periodical data gathering activity that allows collecting several water features' in a database. This periodical activity takes place at a fixed-pace (i.e., $1$ survey per hour), thus granting data availability, reliability, and consistency. 
More in detail, every hour the prototype is woken up to (i) measure the values of interest, (ii) pre-elaborate the data, (iii) prepare the packets to be sent, and finally (iv) transmit them.
Those transmissions are carried out thanks to the Sigfox \ac{BS} that mainly acts as a relay toward the Sigfox Cloud.
The Sigfox Cloud-based core network works to control and manage the \acp{BS} and \ac{IoT} devices. At the same time, it guarantees data connectivity between the \acp{BS} and the Internet, thus allowing gateway functionalities relying on back-hauling systems.
The following logical node involved in data processing and usage is the application server.

The front-end of the WaterS system, i.e., the end-user application, is conceived to monitor the water quality parameters. Indeed, once the required authentication procedure by the end-user to the application server is concluded, the latter is in charge of either authorizing/denying access to information by accepting/rejecting requests.

\textbf{Protocol Compliance and Main Features.} 
Leveraging the compliance to the Sigfox standard, WaterS transmits the sensed data to the Sigfox Base Station. Concerning the main protocol features, every time a message is sent from WaterS to the Sigfox \ac{BS}, it is suddenly transmitted to the Sigfox Cloud infrastructure. The aforementioned communication is, in fact, straightforward, since the Sigfox \ac{BS} mainly act as the reference \ac{RAN} for the Sigfox devices toward the cloud infrastructure.
Once data reach the Sigfox Cloud, this component is in charge of executing specific callbacks so that the information coming from the \ac{IoT} device can be properly stored, leveraging dedicated.

It is worth noting that, once data are in the Sigfox Cloud, and hence made available for the Application Server, the dedicated processing logic is triggered to properly handle the messages and the information within them. Without loss of generality, two custom routines are developed to forward a message received by the Sigfox Cloud to the Application Server. In particular, two different kinds of messages (from now on, also referred to as frames) are defined:
\begin{itemize}
    \item Frame Type $0$, containing measured values coming from the monitoring sensors (temperature, $32$~bits sized float number; pH, $16$~bits long integer number; turbidity, $16$~bits long integer number);
    \item Frame Type $1$, containing information about \ac{GPS} information, such as latitude and longitude, both in the form of $32$~bits sized float numbers.
\end{itemize}

\textbf{Deep Learning applied to WaterS.} The prediction capabilities of WaterS are enabled by a \ac{RNN} namely \ac{LSTM}.
In a nutshell, the amount of data gathered transmitted from the Cloud Sigfox to the Application Server is filtered and processed based on the year the surveys belong to.
From this point on, our work will discuss the details on the leading design criteria and the methodological approach. On top of that, we will present the experiments as well as a thorough analysis of the obtained results.

\section{Proposed Approach}\label{sec:approach}
The problem was modeled as a multivariate time series forecasting with stacked \ac{LSTM} networks.
The \ac{LSTM} network architecture has been selected based on its effectiveness in time series prediction and in learning long-term dependencies \cite{Chen2015_BDC, Fu2016_YAC, Yunpeng2017_WISA}. Their gating mechanism, that controls the information flow in the cells, is able to resolve the vanishing or exploding gradients training problem with common RNN networks \cite{Pascanu2013_ICML}. Evidence has also proved that \ac{LSTM} networks are more effective than the conventional \ac{RNN} \cite{Palangi2016_TASLP} \cite{Palangi2016_TSP}. The stacked \ac{LSTM} structure has been chosen because adding depth is a more efficient approach to extract richer features and increase model capacity \cite{Sadreazami2018_LSC} also providing a type of representational optimization \cite{Pascanu2014_ICLR}.

By following a typical machine learning model training, the network’s hyperparameters have been set, based on evaluation on a validation set, while weights and deviations are updated by using algorithms that minimize a loss function.

When the training process ends, final weights and training data are saved for later use and analysis.
The routine is conceived to periodically train the system when new data comes and to provide the prediction related to the water quality parameters.
Before the training step, investigations about possible correlations among data to evaluate the informativeness of the overall dataset were performed.  To this aim,  the Pearson correlation coefficient was computed by considering all the $3,280$ samples in the dataset. 

\section{Experimental Evaluation}\label{sec:experiments}
In this section, we present the experimental evaluations. We describe the structure of the dataset used in the experiments, the metrics used to evaluate the model, and the experimental settings.

\subsection{Dataset}
This section has been divided into three parts: Dataset Characteristics, Survey  Campaign, and Preprocessing. It is worth remarking that all the available water quality features have been used in the design of the $2$-layers stacked \ac{LSTM} network.
\subsubsection{Dataset Characteristics}
The dataset was originally published as a technical report of the Project \emph{Tiziano}~\cite{progettoTiziano}.
Project Tiziano is an underground waters monitoring system that collects qualitative and quantitative information for the Puglia region, in Italy.
It is worth noting that Project Tiziano's dataset may represent the most extensive dataset regarding the water quality parameter samples collection.
The dataset contains water quality information, collected by means of surveys, each one containing different parameters, such as:
(i) average temperature, (ii) electric conductivity, (iii) medium dissolved oxygen, and (iv) pH. The values have been collected from different station probes, located in the Italian Apulia region.
Data gathering activities were carried out over five years (from $2008$ to $2012$) at a fixed pace, i.e., one sample per hour.

\begin{figure}[htbp]
  \centering
  \includegraphics[angle=0, width=\columnwidth]{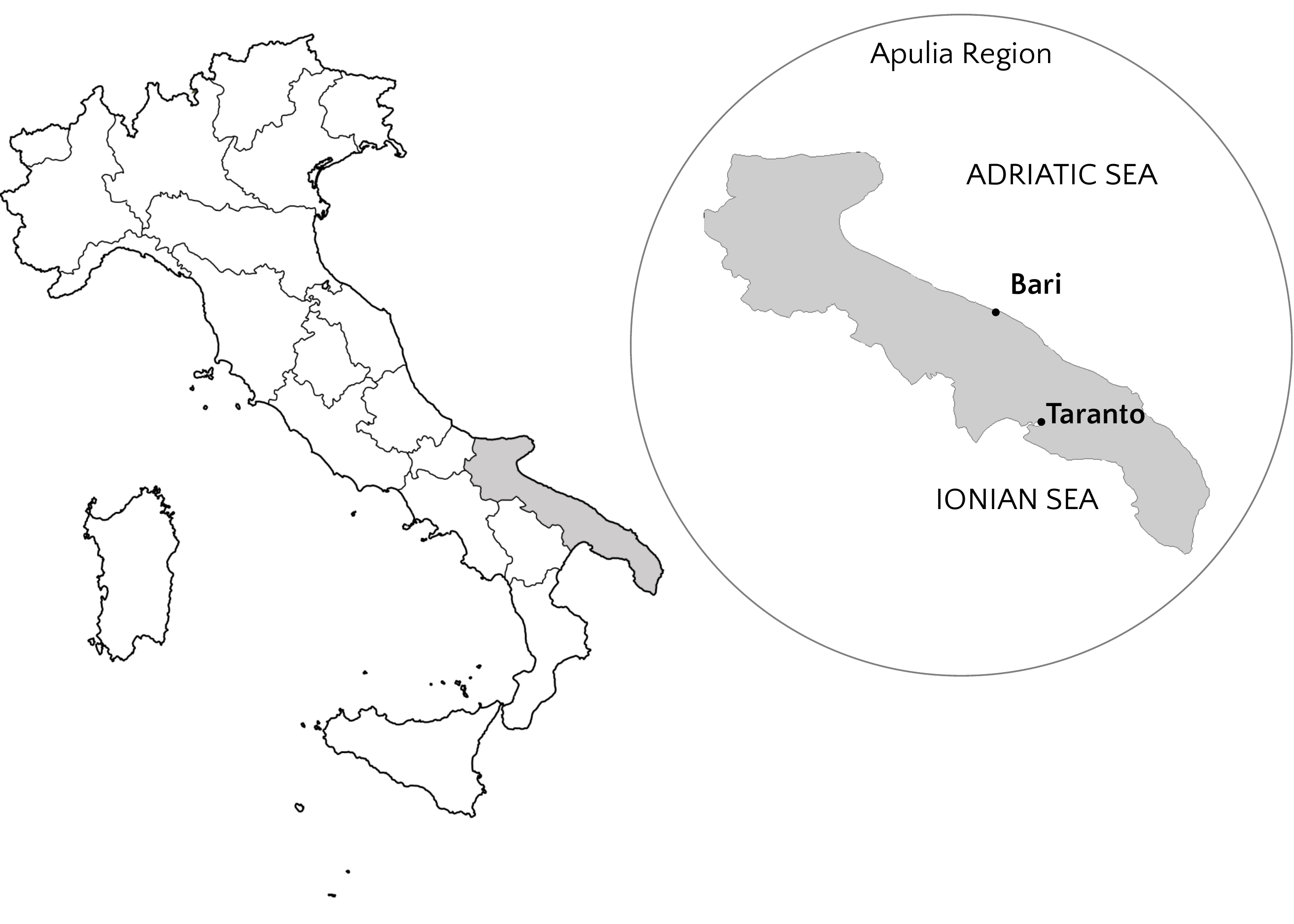}
  \caption{Region of interest for the Tiziano Project and data source identification.}
  \label{fig:apulia_region}
\end{figure}

\subsubsection{Organization Criteria}
The  surveys  were  selected  and  filtered  according  to  the following criteria:
\begin{itemize}
    \item reference place of the probe: city town of Bari. This choice is motivated by the fact that this is the measurement point that is closest to the place in which the WaterS prototype has been tested.
    \item hours of the survey: 9 am, 12 pm, 6 pm. The rationale lies in data availability. A preliminary evaluation during the cleaning phase demonstrated that no relevant improvements in the forecasting activities could be achieved due to limited variations of the surveyed variables.
    \item year for the surveys: from $2009$ to $2011$. This time interval has been verified to be the longest time period available with contiguous information. This choice allowed a speed up during the pre-processing phase.
\end{itemize}
The resulting dataset is composed of $3,280$ samples and is characterized by four different features. 

\subsubsection{Preprocessing}
We have analyzed and filtered the data transmitted from the \textit{Sigfox Cloud} to select the relevant surveys on a per-year basis.
On top of the dataset creation phase, we did the following preliminary procedures:
\begin{enumerate}[(i)]
    \item Data filtering;
    \item Standardization with $z$-score;
    \item Samples organization in the $(x, y)$ form;
    \item Split in \textit{training set}, \textit{validation set}, \textit{test set}.
\end{enumerate}
As an outcome, the \textit{Input Layer} of our \ac{LSTM} neural network (see Figure~\ref{fig:neural_network_waters}) could work on the sensed data.

Data provided in the \emph{Progetto Tiziano} dataset were filtered to remove inaccuracies/inconsistencies on observations and, as previously recalled, to examine the results for a particular time period. Later, the standardization phase has been carried out on the filtered data by means of $z$-score as depicted in the Eq.~\ref{eq:zscore} to provide the same scale to each data point and to allow the comparison with the other (standardized) variables~\cite{Liu2019}.
Further, let $x$ the data sample value, $\mu$ the mean of the training samples and $\sigma$ its standard deviation, the used formula for the standardization phase was:
\begin{align}
\label{eq:zscore}
    z=\frac{x-\mu}{\sigma}
\end{align}

For each feature of the dataset, the resulting distribution has a mean value equal to $0$ and a standard deviation of $1$.

\subsection{Evaluation Metrics}
Between all the evaluation metrics used in machine learning, the classification ones were discarded considering only the typical regression ones. The loss function that minimizes the train fitting process is the \ac{MSE}, as shown in Eq. \eqref{eq:mse}. As a further consideration, smaller values of the \ac{MSE} correspond to a better fitting line. 
The training process involved both \ac{MAE} and \ac{CP}.
On the one hand, the \ac{MAE} defined in Eq.~\eqref{eq:mae} is the average of all absolute errors and it is supposed to measure the accuracy by computing the difference between forecasted and observed value.
On the other hand, the Cosine Proximity as shown in Eq.~\eqref{eq:cosine_prox}, is a measure of the similarity between two vectors. This metric is adopted to compute the cosine proximity between the predicted value and actual value.
It is hereby assumed that $N$ is the number of data samples, $\boldsymbol{y}$ is the actual value for data points, and $\boldsymbol{\hat{y}}$ is the vector denoting the predicted values returned by the model.
\begin{align}
\label{eq:mse}
    MSE &= \frac{1}{N} \sum_{i=1}^{N} (y_i - \hat{y_i})^2 \\
\label{eq:mae}
    MAE &= \frac{1}{N} \sum_{i=1}^{N} |(\hat{y_i} - y_i)| \\
\label{eq:cosine_prox}
    CP &= - \frac{\boldsymbol{y}\cdot\boldsymbol{\hat{y}}}{||\boldsymbol{y}||\cdot||\boldsymbol{\hat{y}}||}
\end{align}
The final metrics values on the test set, used to summarize and assess the quality of this deep learning model, are detailed in Table \ref{tab:metrics}.
\begin{table}[htbp]
    \caption{Classification metrics with standardized data on the test set.}
    \centering
        \begin{tabular}{P{2cm}|P{2cm}|P{2cm}}
         \textbf{Mean Squared Error} & \textbf{Mean Absolute Error} & \textbf{Cosine Proximity} \\ \hline
         $0.09171802$ & $0.20443536$ & $0.93728703$   \\ \hline
        \end{tabular}
    \label{tab:metrics}
\end{table}

The number of units of the hidden and output state of every cell, referring to the number of times that the entire training set was processed by the \ac{LSTM} network, was made dynamic, in order to estimate the optimal training values for the hyperparameters. The maximum values for the number of state units and epochs were set to $100$ and $1000$, respectively.
When designing this kind of network, the challenge is to find out the optimal trade-off in tuning the number of training epochs and the particular parameters of the \ac{LSTM} cells, i.e. the hidden and the output state vector size. This results to be challenging since the number of cells must be kept constant.

\subsection{Experimental Settings}
Similarly to what happens in the classical supervised machine learning problems, the input data for the \ac{LSTM} network must be defined in the $(x, y)$ form, in which $x$ describes a $[1\times4\times3]$ multidimensional input array. In particular, $4$ is referred to the different water quality parameters probed, and $3$ is the number of the different surveys preceding the forecasting value in time series. On the other hand, $y$ is a $[1\times4]$ dimensional output array, with the single survey succeeding the first $3$. In the proposed solution, every $3$ surveys there is a timestamp aggregation, thus allowing the creation of the $x$ input array and binding to it the $y$ output array, which represents the first survey of the dataset following the last timestep in the $x$ input group.

As a design choice, the resulting dataset contains $3$ main parts~\cite{marsland2015, beysolow2017}:
\begin{itemize}
    \item training set: $1640$ samples ($50$\% of the dataset) as the sample data to fit the model;
    \item validation set: $820$ samples ($25$\% of the dataset) as the sample data used for the evaluation of the training results after every epoch;
    \item test set: $820$ samples ($25$\% of the dataset) as the sample data used to evaluate the final performances of the resulting trained \ac{LSTM} network.
\end{itemize}
\begin{figure}[!ht]
    \centering
    \includegraphics[angle=0, width=\columnwidth]{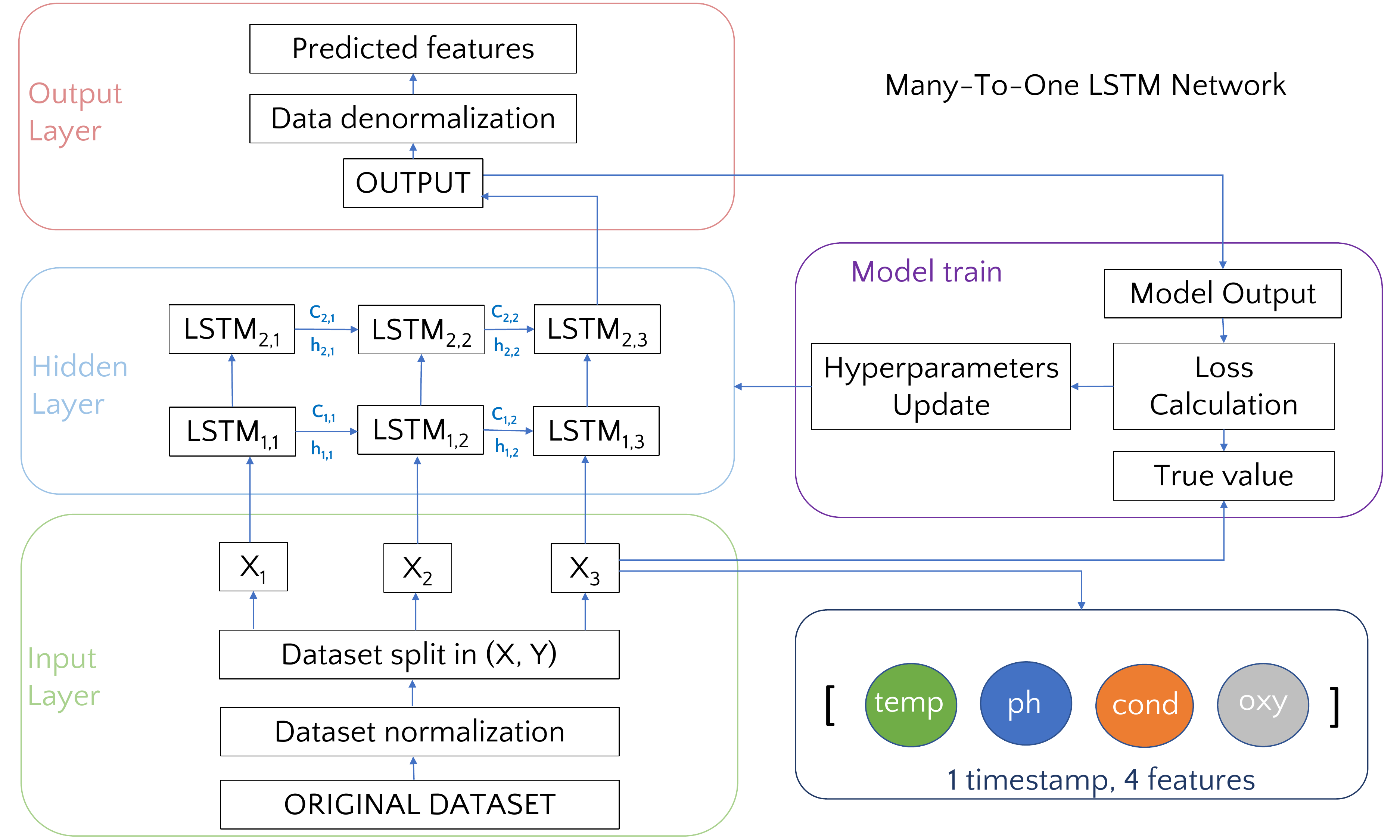}
    \caption{Structured view of the designed Neural Network.}
    \label{fig:neural_network_waters}
\end{figure}

As for the definition of the \ac{LSTM} network as shown in Figure \ref{fig:neural_network_waters}, three different layers were created: (i) the input layer, (ii) the hidden layer, and (iii) the output layer.
In a nutshell, the input layer accepts the $[1\times4\times3]$ multidimensional arrays representing the water quality surveys that precede the forecasting. The hidden layer is designed as the middle layer (i.e., core) of the network, with a $2$-layers stacked \ac{LSTM} configuration and a variable number of units of the hidden state for each cell, according to the training parameters. The output layer returns the results of the network, providing a $1\times4$ multidimensional array with the $4$ different forecasted features.

Both the input and the hidden layers use the \ac{ReLU} functions as the activation function. The model uses the Adam algorithm \cite{kingma2014adam} as an optimization algorithm; the choice is motivated by the fact that this solution is specifically meant for optimization and can be used instead of the classical stochastic gradient descent procedure to update network weights iteratively based on the training data \cite{Zhang2018, SWJ19}.

\section{Results}\label{sec:results}
Table \ref{tab:pearson_correlation} shows the Pearson correlation between the considered parameters. Specifically, water temperature has a negative correlation with conductivity, oxygen, and pH at the significant level of $0.01$. Conductivity has a negative correlation with oxygen, and has a positive correlation with pH at the significant level of $0.01$. Moreover, based on available data, conductivity and pH have shown a positive correlation. Finally, oxygen and pH show the highest correlation value (i.e., a negative correlation of $-0.33209$).
\begin{table}[htbp]
    \caption{Pearson correlation coefficient matrix of water quality parameters.}
    \centering
        \begin{tabular}{P{1.4cm}|P{1.5cm}|P{1.4cm}|P{1.2cm}|P{1.2cm}}
        \textbf{Water \ Parameters} & \textbf{Temperature} & \textbf{Conductivity} & \textbf{Oxygen} & \textbf{pH} \\ \hline
        \textbf{Temperature} & $1.00000$ & $-0.20676$ & $-0.18213$ & $-0.17926$ \\ \hline
        \textbf{Conductivity} & & $1.00000$ & $-0.17549$ & $0.24709$\\ \hline
        \textbf{Oxygen} & & & $1.00000$ & $-0.33209$ \\ \hline
        \textbf{pH} & & & & $1.00000$ \\ \hline
        \end{tabular}
    \label{tab:pearson_correlation}
\end{table}

After the preprocessing phase, the \ac{LSTM} network has been defined and trained in order to predict water parameters.
To verify if the training is consistent and the design phase is completed, we decided to tune the two main hyperparameters, the number of state units of the hidden vector of each \ac{LSTM} cell, and the number of epochs, in order to choose the optimal configuration. At first, training performances were evaluated for every configuration using the validation set and only considering the loss function. The final generalization error was then computed using the 3 main considered metrics on the surveys of the test set. The optimal configuration with the lowest \ac{MSE} and \ac{MAE} and the highest \ac{CP} was found with $70$ state units in the cells of the hidden layer and $40$ training epochs.
\begin{figure}[!ht]
  \centering
  \includegraphics[angle=0, width=\columnwidth]{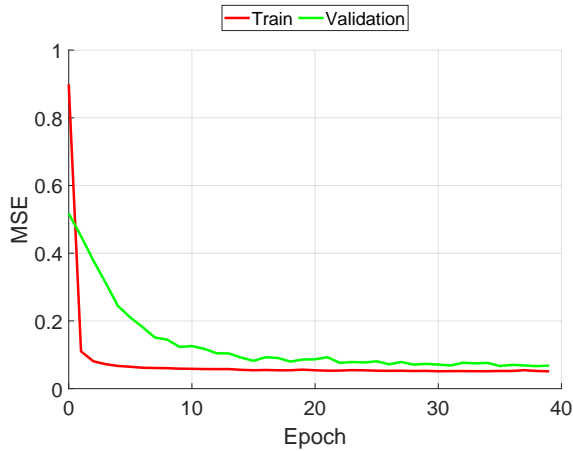}
  \caption{Comparison between the loss of training and validation set, on increasing epochs.}
  \label{fig:train_loss}
\end{figure}

\begin{figure}[htbp]
  \centering
  \includegraphics[angle=0, width=\columnwidth]{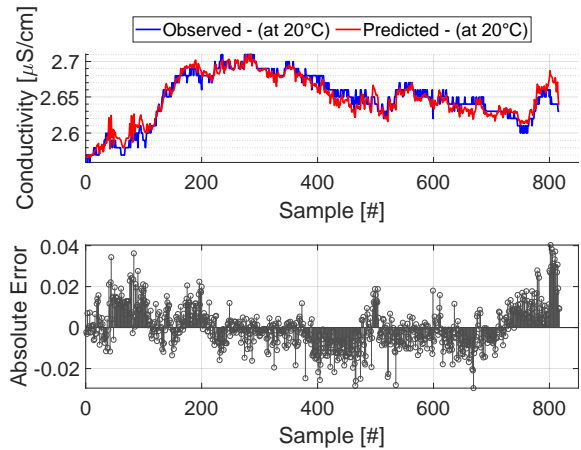}
  \caption{Comparison between the observed and the predicted values of Conductivity.}
  \label{fig:conductivity}
\end{figure}

\begin{figure}[htbp]
  \centering
  \includegraphics[angle=0, width=\columnwidth]{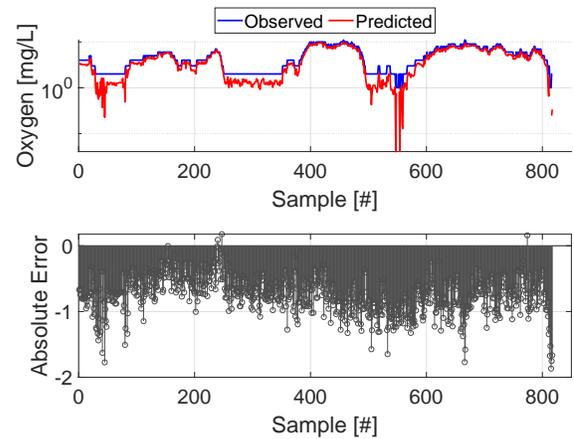}
  \caption{Comparison between the observed and the predicted values of Oxygen dissolved in the water.}
  \label{fig:oxygen}
\end{figure}

Figure \ref{fig:conductivity} shows the values related to conductivity. In particular, here the spread between the values observed and those that have been predicted is pretty clear. The predicted values do not sensibly differ from those that are measured. The absolute values of the error spread from a maximum of $0.04$ to a minimum of $-0.02$.
Figure \ref{fig:oxygen}, instead, shows the amount of oxygen in the water samples. It is worth noting that, even when the data propose significant changes, our \ac{LSTM} solution can properly chase those variations. In this case, the absolute values of the error are always below $0$, with an average value lower than $1$.

\begin{figure}[htbp]
  \centering
  \includegraphics[angle=0, width=\columnwidth]{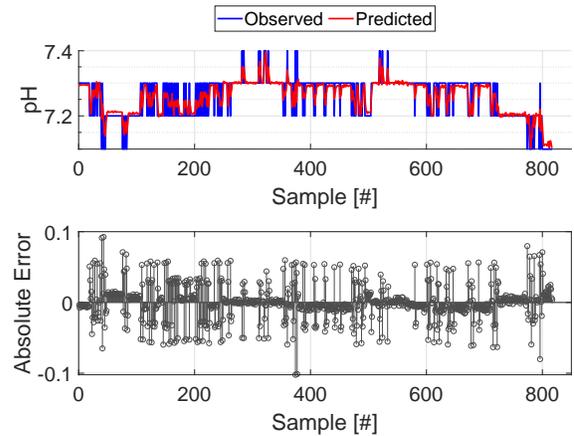}
  \caption{Comparison between the observed and the predicted values of water pH values.}
  \label{fig:ph}
\end{figure}
\begin{figure}[htbp]
  \centering
  \includegraphics[angle=0, width=\columnwidth]{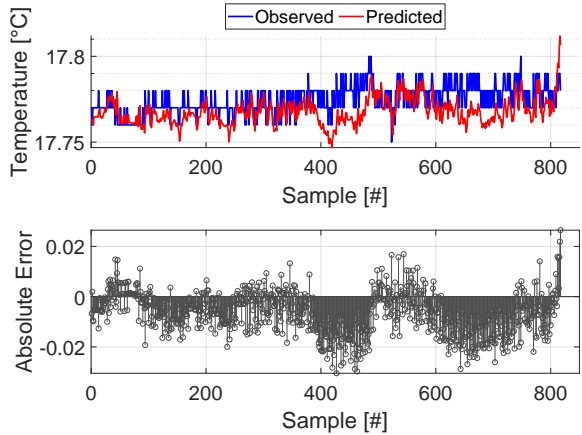}
  \caption{Comparison between the observed and the predicted values of water Temperature values.}
  \label{fig:temperature}
\end{figure}
Figure \ref{fig:ph}, describes both the observed and the predicted values of the pH of the water samples. In this particular case, the spread between the values is not meaningful since the measured ones were always in between $7.4$ and $7$. As a consequence, the prediction is pretty straightforward.
Similarly, as for water temperature (see Figure~\ref{fig:temperature}), the \ac{LSTM} network demonstrated its potential in adapting to the extremely limited variability of a variable of interest. The measured values of the temperature were never higher than $17.9~\si{\degree}C$ and never below $17.75~\si{\degree}C$. As a consequence, the absolute error was always extremely limited.

Since the proposed \ac{LSTM} solution is complex and extremely demanding in terms of computational resources, the whole evaluation and procedures were conducted on a dedicated cloud server with an Intel Xeon CPU W3520, $16$~GB of RAM and a dedicated nVidia \ac{GPU} GeForce GTX 1060 was used, along with an Ubuntu Linux 18.04 LTS operating system properly configured with Keras~\cite{gulli2017deep} and Tensorflow~\cite{abadi2016tensorflow}.

\section{Discussion and Further Directions}\label{sec:disc-way-forw}
\textbf{System Integration.}
The present contribution focused on the applied deep learning to an \ac{IoT} solution for handling both the data gathering phase and the analysis of water quality parameters. We notice that the reference analysis is the result of a multi-varied time series process.
Since the WaterS \ac{IoT} device has been conceived and developed as an embedded microcontroller-based solution, it is worth investigating the feasibility of the integration of the proposed recurrent neural network to the involved device.
In particular, the integration of the prediction routines on a Microchip ARM-Based SAMD21 results to be challenging due the $32$~bit \ac{MCU} with a $256$~kB Flash Memory onboard and $32$~kB of SRAM.
Indeed, to lower the occupied memory, while minimizing the amount of energy required for all the operations, the optimization of the source code has been carried out in terms of both \ac{ROM} memory, with a $4$~kB footprint, thus occupying about $16$\% of the device's ROM and $12$~kB, i.e, $37.5$\%, of the available SRAM.

Given the extremely constrained computational capabilities of the \ac{IoT} device, WaterS could be used to leverage the already-trained neural network instead of autonomously providing this result. Since this procedure generates a reference data structure of $400$~kB large, which exceeds the onboard Flash by $156$\%, the \ac{IoT} device could still support it downstream of dedicated hardware modification.

Although deep learning has been proved suitable for the reference application, its implementation in the current WaterS \ac{IoT} end-node is not straightforward.
This is an extremely demanding routine, from the computational point of view. A similar consideration can be applied to the energy supply, which will be extremely over-stressed during the training phase. This aspect may lead to the unfeasibility of the implementation of the designed routines on top of constrained \ac{IoT} devices.
As a matter of fact, the WaterS end-node cannot be completely autonomous in characterizing the neural network and in estimating the associated weights. Nevertheless, the Waters architecture could still take advantage of deep learning solutions integrating the \ac{LSTM} network within the powerful nodes in the Sigfox network (e.g., the \acp{BS}).

\textbf{Protocol Suitability.}
At the time of writing, our solution adopts the Sigfox protocol only for uplink communications. Given the compliance of the WaterS architecture to the standard Sigfox, it could be feasible to remotely transmit the \ac{LSTM} weights in order to update them onboard of the \ac{IoT} device. Indeed, leveraging the downlink capabilities, incremental weights updates could be received by the device, thus realizing an \ac{OTA} procedure for improving processing. Further, as future work the conceived solution could also be cross-compared with reference to \ac{LPWAN} technologies such as LoRa and NB-IoT.
\begin{figure}[htbp]
  \centering
  \includegraphics[angle=0, width=\columnwidth]{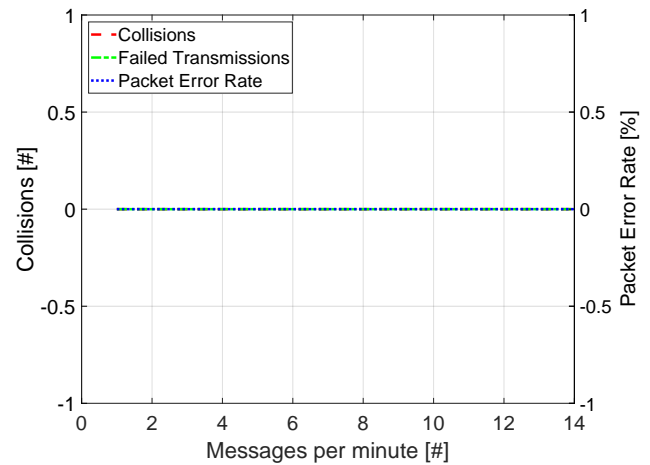}
  \caption{Network performance with 14 Sigfox-compliant devices.}
  \label{fig:14_devs}
\end{figure}
\begin{figure}[htbp]
  \centering
  \includegraphics[angle=0, width=\columnwidth]{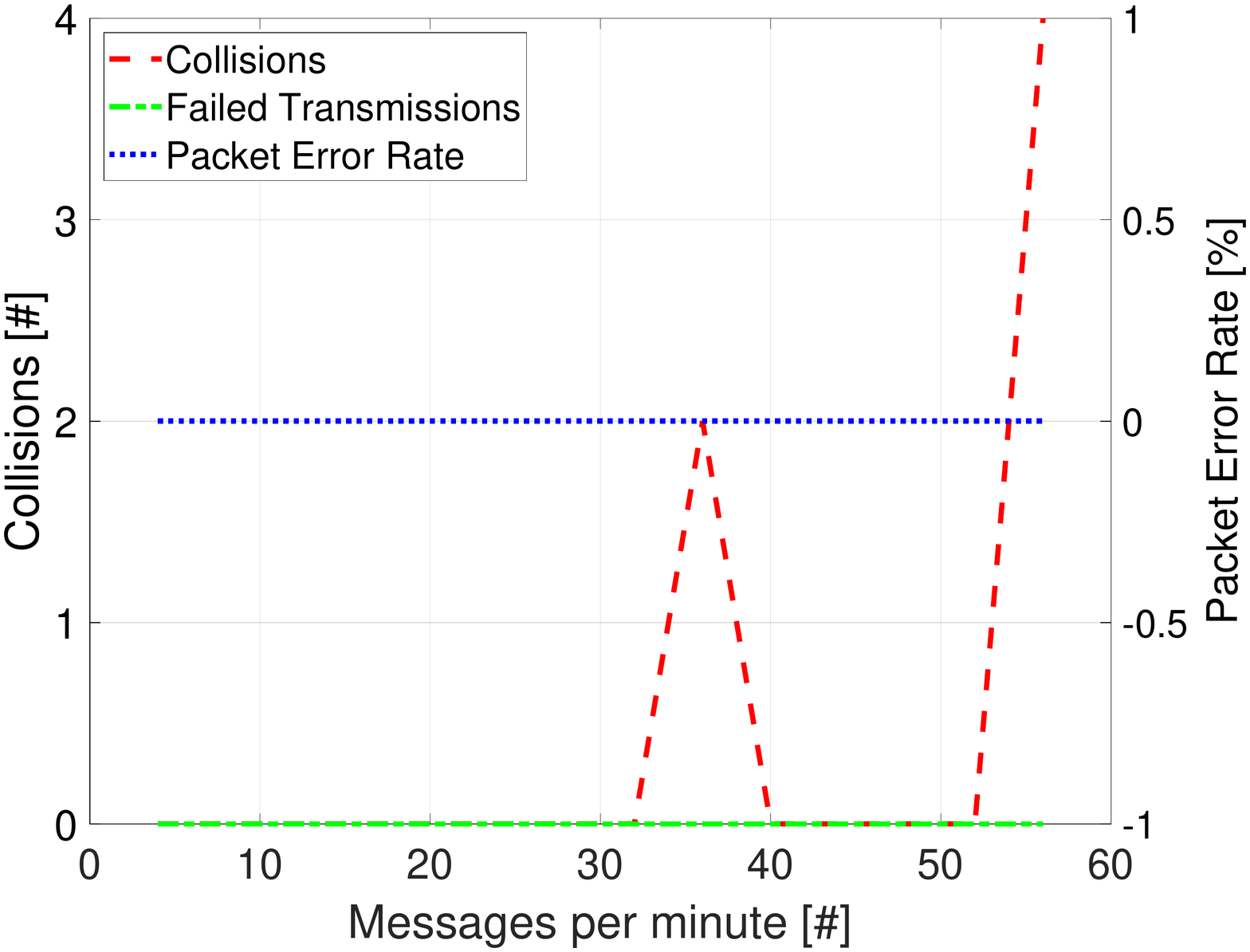}
  \caption{Network performance with 56 Sigfox-compliant devices.}
  \label{fig:56_devs}
\end{figure}
\begin{figure}[htbp]
  \centering
  \includegraphics[angle=0,width=\columnwidth]{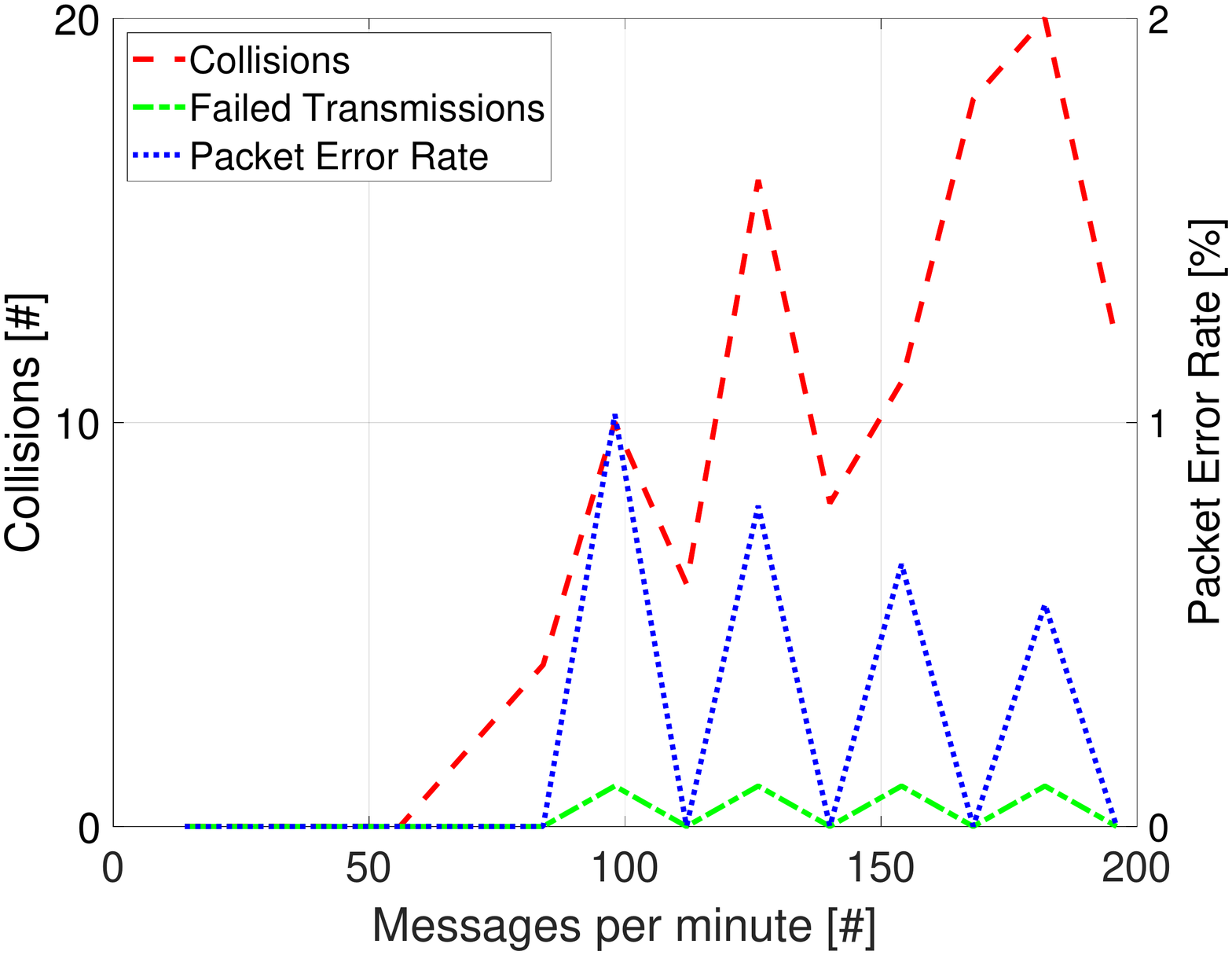}
  \caption{Network performance with 200 Sigfox-compliant devices.}
  \label{fig:200_devs}
\end{figure}
\begin{figure}[htbp]
  \centering
  \includegraphics[angle=0,width=\columnwidth]{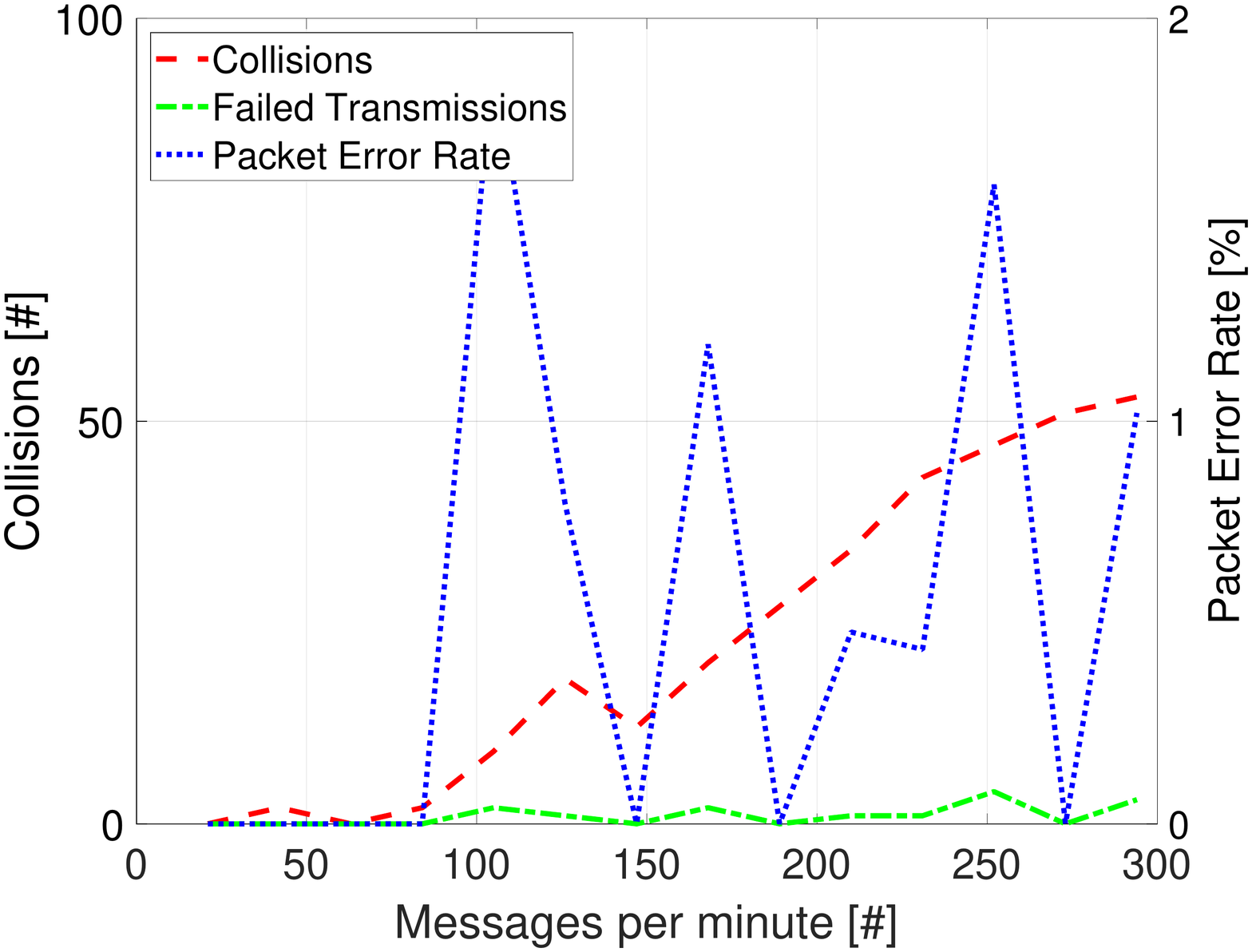}
  \caption{Network performance with 300 Sigfox-compliant devices.}
  \label{fig:300_devs}
\end{figure}
\begin{figure}[htbp]
  \centering
  \includegraphics[angle=0,width=\columnwidth]{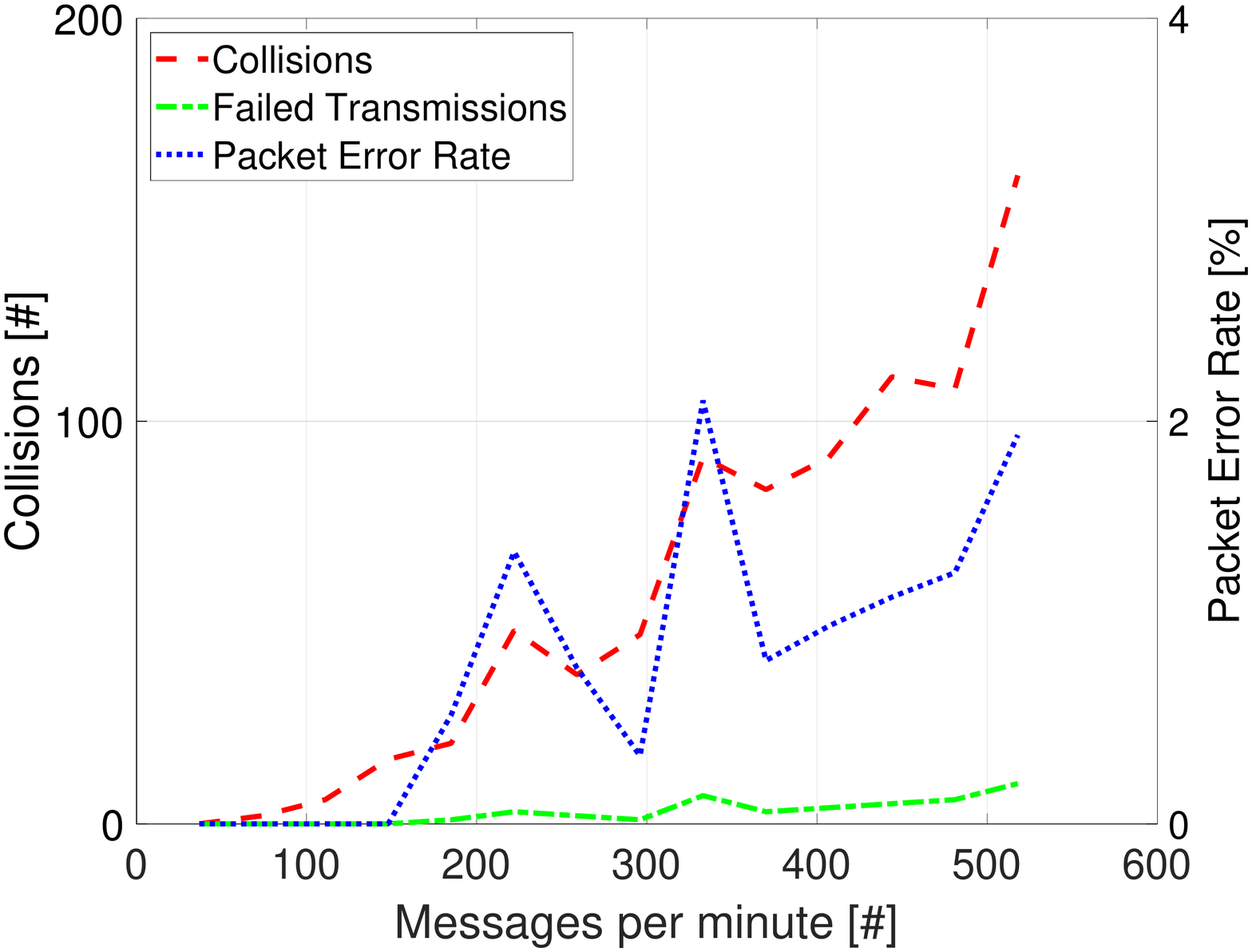}
  \caption{Network performance with 520 Sigfox-compliant devices.}
  \label{fig:520_devs}
\end{figure}
\begin{figure}[htbp]
  \centering
  \includegraphics[angle=0, width=\columnwidth]{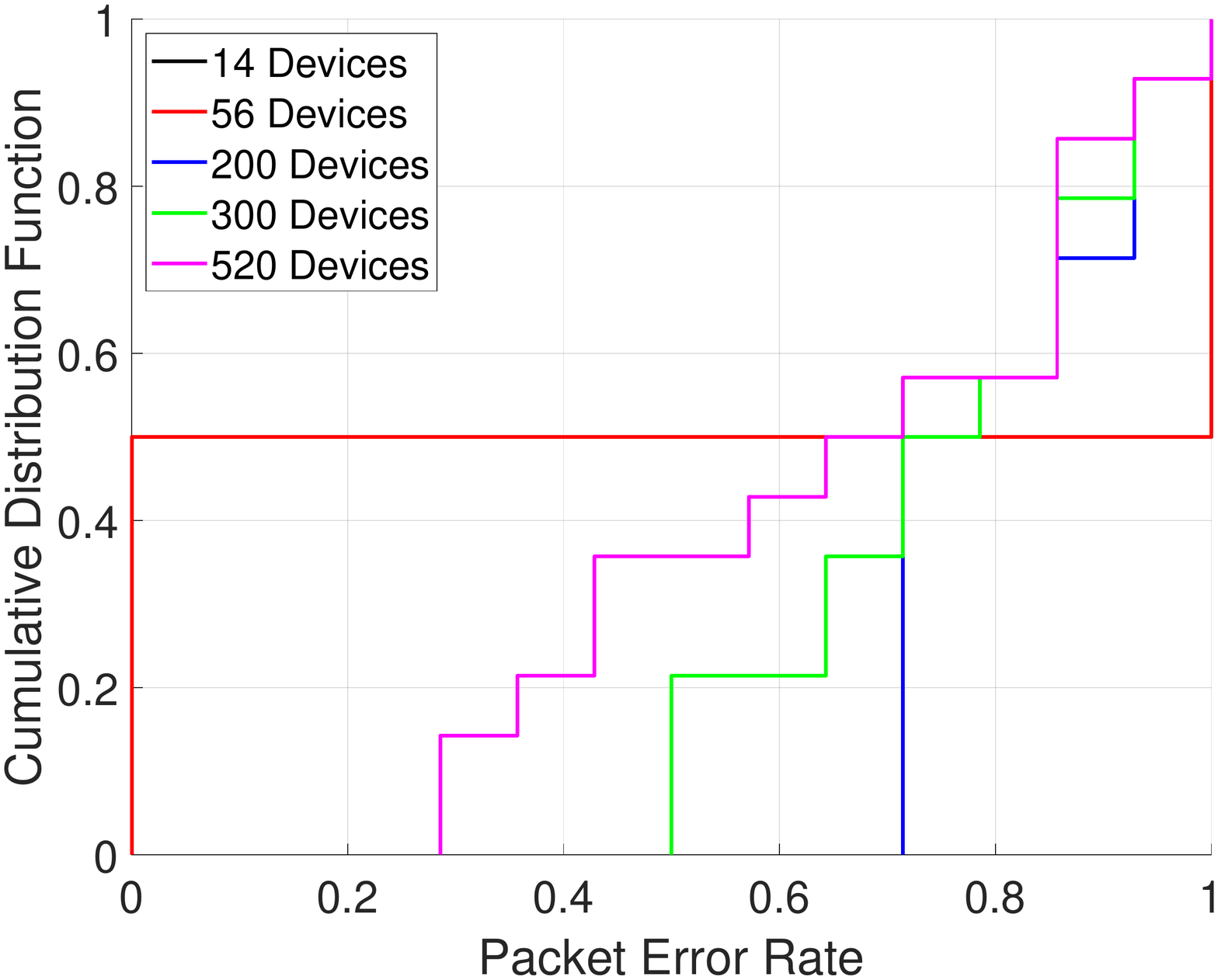}
  \caption{\acl{CDF} for the five network configurations.}
  \label{fig:cdf}
\end{figure}

\textbf{Considerations on Scalability.}
The WaterS architecture is conceived as modular and one of the most promising exploitation perspectives is related to the increase of the number of \ac{IoT} devices. To provide further insight into this assessment, the network traffic must be preliminarily estimated.
To address the optimal communication parameters (i.e., number of timeslots used, number of transmitted packets) while testing large-scale deployments, the network configurations have been supposed to be deployed with $14$, $56$, $200$, $300$, and $520$ Waters end-devices. The latter value represents an upper bound to the reference Tiziano Project that involved a total number of $393$ sensing units.

The results shown in Figures from \ref{fig:14_devs} to \ref{fig:520_devs} describe the trends of the \ac{PER}, together with the number of packets lost in transmissions.
Overall, with an increasing number of end-devices, the \ac{PER} increases as well up to $4$\% with the largest envisioned deployment.
 
In detail, with $14$ end-devices involved, the number of collisions is close to $0$, except for some cases which, however, do not significantly affect the overall percentage. Therefore, it can be concluded that with a limited number of devices, the losses are negligible. Similar considerations can be made for a network made up of $56$ WaterS devices (shown in Figure~\ref{fig:56_devs}).
With $200$ end-devices, instead, (see Figure \ref{fig:200_devs}), the total number of lost packets is slightly less than $30$ and the \ac{PER} has an extremely limited impact, as low as $1$\%. This value is increased by $50$\% when the number of devices becomes $300$ units, as shown in Figure \ref{fig:300_devs}, in which case the number of lost packets approaches $60$.
In Figure~\ref{fig:520_devs}, with $520$ end-devices, it is worth noting that the error percentage never exceeds $5$\%, also taking, in this case, the overall \ac{PER} below acceptable thresholds.

Lastly, in Figure~\ref{fig:cdf}, the \ac{CDF} values of \ac{PER} are shown for the proposed network configurations. Between the $70$-th and the $80$-th percentile, it is shown that the increasing number of devices leads to an increase in the \ac{PER}. With a number of $14$, $56$, and $200$ devices, in fact, the $70$-th percentile is reached with \ac{PER} equal to $90$\%. With larger deployments, e.g., $300$ or $520$ devices, the $80$-th percentile is reached for values that are greater than $80$\%.

\section{Conclusion}\label{sec:conclusion}
This work presented the WaterS architecture, an \ac{IoT} solution that leverages the water monitoring capabilities, and the compliance to one of the most promising candidates in the context of \acp{LPWAN}. Leveraging its prototypical nature, the WaterS system has been proposed for an enhancement based on the employment of neural network solutions. The idea has been proven of interest since a neural network can be used to process gathered data and promote water quality analysis. The \ac{LSTM} applied to our ecosystem achieves a \ac{MAE} as low as $~0.20$, a \ac{MSE} of $~0.092$ and a \ac{CP} equal to $~0.94$.
Further, this work demonstrated an interesting networking perspective, since an increasing number of Sigfox-enabling end-devices may lead to \ac{PER} values as low as $4$\% in the largest envisioned deployment, which includes more than $500$ end devices.
Even though the results are noticeable, in the near future, WaterS could be adopted as a system to identify potential waters' anomalies.
One of the most thrilling research perspectives is the development of a federated machine learning approach. This approach could sensibly improve data analysis in case of missing data/surveys or transmission losses and errors while preserving forecasting capabilities.
Finally, the source code of WaterS has been released as open-source. This enables the research community to verify our claims, and use the released code as a ready-to-use basis for further protocol improvement and comparison.

Although the proposed results are of relevance, it is worth specifying that the analysis could be strengthened by including the depth of the probing point as a parameter. In fact, from a hydrogeological point of view, such detail on well waters represent an important variable in establishing correlations between variables, since the electrical conductivity and the salinity are strictly bounded one with the other. In terms of the neural network, this aspect could be dealt with by including some non-linear or, in general, more complex correlation functions.
Therefore, the proposed results must be considered as an interesting preliminary result, which, at the same time, demonstrates that the applicability of the method must be extended to wider information frameworks.
Lastly, the reported results are obtained on a dataset referred to as coastal waters. It would be of importance to investigate if such a solution may be applied to a continental aquifer, as well as the quality of the obtained results.

\section*{Acknowledgments}
The authors would like to thank Prof. D. Fidelibus, Prof. T. Di Noia, C. Pomo, and W. Anelli for their contributions and support to this work. 

\balance
\bibliographystyle{IEEEtran}
\bibliography{waters}

\end{document}